\documentclass[pra,aps,singlecolumn,amsmath,showpacs,floatfix]{revtex4}
\usepackage{tipa}
\usepackage{epsfig}
\usepackage{amsmath}
\usepackage{mathrsfs}
\usepackage{float}
\usepackage{color}

\allowdisplaybreaks
\begin{document}

\title{Separability criterion for bipartite states and its generalization to multipartite systems}

\author{Jie-Hui Huang$^{1,*}$, Li-Yun Hu$^{1}$, Lei Wang$^{2}$, and Shi-Yao Zhu$^{3}$}

\affiliation{$^1$College of Physics and Communication Electronics, Jiangxi Normal University, Nanchang 330022, People's Republic of China\\
$^2$College of Physics, Jilin University, Changchun 130021, People's Republic of China\\
$^3$Beijing Computational Science Research Center, Beijing 100084, People's Republic of China}

\begin{abstract}
A group of symmetric operators are introduced to carry out the separability criterion for bipartite and multipartite quantum states. Every symmetric operator, represented by a symmetric matrix with only two nonzero elements, and their arbitrary linear combinations are found to be entanglement
witnesses. By using these symmetric operators, Wootters' separability criterion for two-qubit states can be generalized to bipartite and multipartite systems in arbitrary dimensions.
\end{abstract}


\maketitle

Quantum entanglement, introduced in the early days of quantum theory
\cite{EPR&Schrodinger}, not only plays a vital role in
differentiating quantum mechanics from classical mechanics, but also
acts as key resource in quantum information processing. Thus the
verification of quantum entanglement is
fundamentally important in quantum information science. Since the
mathematical description on quantum entanglement was first introduced
by Werner in 1989 \cite{werner}, many entanglement measures have
been proposed, such as distillable entanglement, entanglement cost
\cite{distillation&cost}, relative entropy of entanglement
\cite{entropy}, Schmidt number \cite{Schmidt}, concurrence
\cite{concurrence0,concurrence}, negativity \cite{negativity}, and
so on \cite{horodecki}. Recently, the verification of multipartite entanglement and relevant issues are attracting more and more attention
\cite{residual,ME} owing to the booming interest on quantum many-body systems \cite{manybody}.

The convexity requirement for quantum entanglement, which is in accord with the fact local operations with classical communication (LOCC) can not create or increase quantum entanglement \cite{distillation&cost}, makes it very hard to verify the entanglement of mixed quantum states. Historically, positive
partial transpose (PPT) criterion, proposed by Peres in 1996
\cite{PPTA}, provides a necessary and sufficient condition for
verifying the separability of $2\otimes2$ and $2\otimes3$ bipartite
systems \cite{PPTB}. But for higher dimensional systems, how to distinguish entangled states from separable ones is still a challenging question. In this Letter, we introduce a group of symmetric operators to answer this question. It is shown that all these symmetric operators and their arbitrary linear combinations are entanglement witnesses. We use these symmetric operators to generalize Wootters' concurrence \cite{concurrence0,concurrence} to high dimensional bipartite \cite{generalization,Iconcurrence} and multipartite systems \cite{multiconcurrence}. More importantly, based on these symmetric operators the separability criterion for bipartite and multipartite states is presented in arbitrary dimensions, which is a generalization of Wootters' separability criterion from the two-qubit case to a general case.

In the simplest two-qubit system, a pure state
$|\psi>=\sum_{i=1}^2\sum_{j=1}^2a_{ij}|i_1j_2>$, where the subscripts
\lq\lq $1$\rq\rq ~and \lq\lq $2$\rq\rq ~in the state vectors denote the two subsystems, is separable when it is a tensor
product of a pure state in subsystem \lq\lq $1$\rq\rq ~ and a pure state in
subsystem \lq\lq $2$\rq\rq ~, thus the four coefficients $a_{ij}~(i,j\in \{1,2\})$
have to satisfy the condition $a_{11}a_{22}=a_{12}a_{21}$. In
Ref.\cite{concurrence}, Wootters introduced a \lq\lq spin-flip\rq\rq
~operator $\sigma_y$ (Pauli matrix) in the standard basis to verify
the separability of a two-qubit pure state. That is, a two-qubit
pure state $|\psi>$ is separable iff
$<\psi^{*}|\sigma_y\otimes\sigma_y|\psi>=0$. In the following, we
call the state vector $|\psi>$ and its complex conjugate $|\psi^{*}>$ orthogonal with respect
to the operator $O$, or simply $O$-orthogonal, if they satisfy the
relation $<\psi^{*}|O|\psi>=0$. Wootters' separability criterion for a
two-qubit pure state can then be stated as, it should be
$O$-orthogonal to its complex conjugate with the operator
$O=\sigma_y\otimes\sigma_y$. Here the antisymmetry of the \lq\lq
spin-flip\rq\rq ~operator $\sigma_y$, i.e. $\sigma_y^T=-\sigma_y$
(the superscript \lq\lq T\rq\rq ~stands for matrix transpose
hereafter), plays the essential role in verifying the separability of
a two-qubit state. This $O$-orthogonal relation between a
separable two-qubit pure state and its complex conjugate is violated
by all entangled two-qubit states.

Now we consider a general bipartite case, with the two subsystems having dimensions $D_1$ and $D_2$, respectively. If a bipartite pure state
$|\psi>=\sum_{i=1}^{D_1}\sum_{j=1}^{D_2}a_{ij}|i_1j_2>$ is
separable, the four coefficients $a_{ij}$, $a_{ij^{\prime}}$,
$a_{i^{\prime}j}$, and $a_{i^{\prime}j^{\prime}}$ with
$i,i^{\prime}\in \{1,2,\cdots, D_1\}$ and
$j,j^{\prime}\in\{1,2,\dots, D_2\}$, have to satisfy the condition
$a_{ij}a_{i^{\prime}j^{\prime}}=a_{ij^{\prime}}a_{i^{\prime}j}$, so that this bipartite pure state can be formulated as a tensor product.
This condition, similar to the above two-qubit case, is equivalent
to the following $O^{(i,i^{\prime}:j,j^{\prime})}$-orthogonal
relation,
\begin{subequations} \label{eq1}
\begin{align} \label{eq1a}
<\psi^{*}|O^{(i,i^{\prime}:j,j^{\prime})}|\psi>=0,
\end{align}
\text{with}
\begin{align} \label{eq1b}
O^{(i,i^{\prime}:j,j^{\prime})}=[\sigma_1^{(i,i^{\prime})}-(\sigma_1^{(i,i^{\prime})})^T]\otimes
[\sigma_2^{(j,j^{\prime})}-(\sigma_2^{(j,j^{\prime})})^T].
\end{align}
\end{subequations}
Here the operator $\sigma_1^{(i,i^{\prime})}$ ($\sigma_2^{(j,j^{\prime})}$) in the subsystem \lq\lq $1$\rq\rq ~(\lq\lq $2$\rq\rq) has only one nonzero element \lq\lq 1\rq\rq ~located at row $i$ ($j$) and column $i^{\prime}$ ($j^{\prime}$) in the standard basis,
no matter what dimensions it has. The two antisymmetric (or skew-symmetric) operators,
$[\sigma_1^{(i,i^{\prime})}-(\sigma_1^{(i,i^{\prime})})^T]$ and $[\sigma_2^{(j,j^{\prime})}-(\sigma_2^{(j,j^{\prime})})^T]$, which play the same role as the antisymmetric operator $\sigma_y$ in the two-qubit case, establish a relation among the four coefficients $a_{ij}$, $a_{ij^{\prime}}$,
$a_{i^{\prime}j}$, and $a_{i^{\prime}j^{\prime}}$.

Since the superscript numbers $i,i^{\prime}\in \{1,2,\cdots, D_1\}$
and $j,j^{\prime}\in\{1,2,\dots, D_2\}$ in the above Eq.(\ref{eq1})
can be chosen arbitrarily, the
$O^{(i,i^{\prime}:j,j^{\prime})}$-orthogonal relation (\ref{eq1a}) for a separable bipartite pure state, holds valid for all possible
operators $O^{(i,i^{\prime}:j,j^{\prime})}$ in the form
(\ref{eq1b}). Owing to the linear dependence of the relation (\ref{eq1a}) on the operators $O^{(i,i^{\prime}:j,j^{\prime})}$, we even can replace the operators
$\sigma_1^{(i,i^{\prime})}$ and $\sigma_2^{(j,j^{\prime})}$ by a
random matrix $\sigma_1^{\text{(rdm)}}$ in subsystem \lq\lq $1$\rq\rq ~ and a
random matrix $\sigma_2^{\text{(rdm)}}$ in subsystem \lq\lq $2$\rq\rq ~ to
construct,
\begin{align} \label{eq1add}
O^{\text{(Semi-rdm)}}=[\sigma_1^{\text{(rdm)}}-(\sigma_1^{\text{(rdm)}})^T]\otimes
[\sigma_2^{\text{(rdm)}}-(\sigma_2^{\text{(rdm)}})^T],
\end{align}
and the $O^{\text{(Semi-rdm)}}$-orthogonal relation,
$<\psi^{*}|O^{\text{(Semi-rdm)}}|\psi>=0$, still holds
true if only the pure state $|\psi>$ is separable. Thus we can
conclude that a bipartite pure state in arbitrary dimensions is
separable iff it and its complex conjugate are orthogonal with
respect to all symmetric operators formulated as a tensor product
of two antisymmetric operators in the two subsystems. To verify the separability of a given
bipartite pure state, we only need to pick out a finite number of
linearly independent symmetric operators and check the above
$O^{\text{(Semi-rdm)}}$-orthogonal relation. A violation of the $O^{\text{(Semi-rdm)}}$-orthogonal relation by any
operator $O^{\text{(Semi-rdm)}}$ is enough to declare the existence
of quantum entanglement.

Now we use a complete set of linearly independent symmetric
operators $O^{(i,i^{\prime}:j,j^{\prime})}$, which are directly
constructed by $\frac{1}{2}D_1(D_1-1)$ linearly independent
antisymmetric operators
$[\sigma_1^{(i,i^{\prime})}-(\sigma_1^{(i,i^{\prime})})^T]$ in
subsystem \lq\lq $1$\rq\rq ~ and $\frac{1}{2}D_2(D_2-1)$ linearly independent
antisymmetric operators
$[\sigma_2^{(j,j^{\prime})}-(\sigma_2^{(j,j^{\prime})})^T]$ in
subsystem \lq\lq $2$\rq\rq ~, to define the entanglement measure of a bipartite
pure state $|\psi>$,
\begin{align} \label{eq2}
C^{(2)}(|\psi>)=\sqrt{\sum_{i=1}^{D_1-1}\sum_{i^{\prime}=i+1}^{D_1}\sum_{j=1}^{D_2-1}\sum_{j^{\prime}=j+1}^{D_2}|<\psi^{*}|[\sigma_1^{(i,i^{\prime})}-(\sigma_1^{(i,i^{\prime})})^T]\otimes
[\sigma_2^{(j,j^{\prime})}-(\sigma_2^{(j,j^{\prime})})^T]|\psi>|^2},
\end{align}
where the superscript \lq\lq $(2)$\rq\rq ~denotes the bipartite
system. This operator-based entanglement measure for bipartite pure
states, which we call \emph{O concurrence} in the following, turns
back to Wootters' concurrence for two-qubit systems, and satisfies
the basic requirements for a good entanglement measure. For example,
(i) it presents zero result for all tensor product states, and
positive results for entangled states; (ii) it remains invariant
under all local unitary transformations, and so on. In fact, this
\emph{O concurrence} is the same as measured in the following way,
\begin{align} \label{eq3}
C^{(2)}(|\psi>)=\sqrt{2-\text{tr}(\rho_1^2)-\text{tr}(\rho_2^2)},
\end{align}
with $\rho_{1}$ ($\rho_{2}$) being the reduced density matrix of
subsystem \lq\lq $1$\rq\rq ~ (\lq\lq $2$\rq\rq ), and \lq\lq $\text{tr}( )$\rq\rq ~denoting
matrix trace. Since the equality
$\text{tr}(\rho_1^2)=\text{tr}(\rho_2^2)$ holds for bipartite pure
states, our entanglement measure is equivalent to the entanglement
measure proposed in Ref.\cite{Iconcurrence}, which is called \emph{I
concurrence} owing to its connection to the universal inverter. It
is very interesting that two measures present the same result for
the bipartite entanglement, though they are defined in two different
ways and based on different ideas.

This \emph{O concurrence} can also be extended by convex roof to
measure the entanglement of mixed bipartite states. At present, we
only concern the separability condition of mixed bipartite states,
rather than the evaluation of their entanglement. According to
Werner's criterion \cite{werner}, a mixed bipartite state
$\rho$ is separable only when it can be written as a mixture of
separable pure states,
\begin{align} \label{eq4}
\rho=\sum_{i,j}P_{ij}|\psi_1^i><\psi_1^i|\otimes|\psi_2^j><\psi_2^j|,
\end{align}
with the probabilities $P_{ij}\ge0~\text{and}~\sum_{i,j}P_{ij}=1$. However,
every mixed quantum state has countless types of pure-state
decomposition. Given a density matrix, it is usually very difficult
to prove the existence or nonexistence of a decomposition composed
of product states. In other words, we can easily construct a
separable mixed quantum state by mixing a group of separable states
in the above way (\ref{eq4}). However, it is very difficult to
recover this group of separable states, provided only the density
matrix.

We begin our investigation on the separability of a mixed $D_1\otimes
D_2$ state with a Hermitian matrix $\sqrt{\rho}$, which is the
square root of the density matrix $\rho$. Every pure-state
decomposition of the density matrix $\rho$ is connected to a
$D_{12}\times M$ ($D_{12}=D_1\times D_2\leq M$) right-unitary
transformation $U$ through the relation, $W=\sqrt{\rho}U$. Each
column vector of the matrix $W$, the $i$th column vector denoted as
$|W_i>$ hereafter, is a \lq\lq subnormalized\rq\rq ~pure state in the
$U$-decomposition, and the original density matrix can then be
written as $\rho=\sum_{i=1}^{M}|W_i><W_i|$. Now we suppose that
the density matrix $\rho$ represents a separable mixed state,
which means there exists at least one $U$-decomposition so that all
pure states in this decomposition are tensor product states. That is
to say, every column vector $|W_i>$ is
$O^{\text{(Semi-rdm)}}$-orthogonal to its complex conjugate, with
the operator $O^{\text{(Semi-rdm)}}$ defined in Eq.(\ref{eq1add}).

Since all column vectors $|W_i>$ satisfy the
$O^{\text{(Semi-rdm)}}$-orthogonal relation,
$<W_i^{*}|O^{\text{(Semi-rdm)}}|W_i>=0$, for a particular
$U$-decomposition of the separable mixed state $\rho$, the
symmetric matrix $W^TO^{\text{(Semi-rdm)}}W$ is a hollow matrix
whose diagonal elements are all equal to zero. Equivalently
speaking, a mixed bipartite state $\rho$ is separable, iff
there exists a right-unitary transformation $U$, so that the
symmetric matrix $U^TSU$, with
\begin{align} \label{eq6}
S=(\sqrt{\rho})^TO^{\text{(Semi-rdm)}}\sqrt{\rho},
\end{align}
is a hollow matrix for all operators $O^{\text{(Semi-rdm)}}$ in the
form (\ref{eq1add}). Here we emphasize the above symmetric matrix
$S$ depends only on the given density matrix and an operator
constructed by two random matrices. Now a new question arises, given
a symmetric matrix $S$, whether there exists such a right-unitary
transformation $U$ so that $U^TSU$ is a hollow matrix?

We can find out the answer by using Wootters' method in
Ref.\cite{concurrence}. A symmetric matrix $S$ can be transformed to
a hollow matrix in the way of $U^TSU$, only when its maximal
singular value is no larger than the sum of the rest singular
values. Supposing $\{\lambda_i\}$ are singular values of the
symmetric matrix $S$ in decreasing order, the above statement is
equivalent to the following condition,
\begin{align} \label{eq7}
\lambda_1\leq\sum_{i=2}\lambda_i.
\end{align}

On the contrary, if the above condition (\ref{eq7}) is not satisfied
for even one particular operator in the form (\ref{eq1add}), e.g.
$O^{\text{(particular)}}=\sum_{i=1}^{D_1-1}\sum_{i^{\prime}=i+1}^{D_1}\sum_{j=1}^{D_2-1}\sum_{j^{\prime}=j+1}^{D_2}c_{i,i^{\prime}:j,j^{\prime}}O^{(i,i^{\prime}:j,j^{\prime})}$,
the mixed state $\rho$ must be entangled. Please see
Eq.(\ref{eq1b}) for the definition of the operator
$O^{(i,i^{\prime}:j,j^{\prime})}$. Without loss of generality, we
assume the first coefficients $c_{12:12}$ has the maximal modulus
among all the complex coefficients
$\{c_{i,i^{\prime}:j,j^{\prime}}\}$. In this case, the average
\emph{O concurrence} of the pure states, i.e. column vectors
$\{|W_k>\}$, in an arbitrary $U$-decomposition,
$C^{(2)}_U(\rho)=\sum_{k=1}^{M}C^{(2)}(|W_{k}>)$, is larger
than zero, because
$C^{(2)}_U(\rho)\geq\frac{1}{\sqrt{|c_{12:12}|}}(\lambda_1-\sum_{j=2}^{D_{12}}\lambda_j)$.
So, given a mixed bipartite state $\rho$,
if the condition (\ref{eq7}) is violated by even one operator in the
form (\ref{eq1add}), this mixed state is an entangled state. One of
main conclusions in this Letters can now be concluded, a mixed
bipartite state in arbitrary dimensions is separable only when the
condition (\ref{eq7}) for the singular values of the symmetric
matrix (\ref{eq6}) is generally true for \emph{all} operators
$O^{\text{(Semi-rdm)}}$ in the form (\ref{eq1add}). This is a
necessary and sufficient condition. On the one hand, if the
condition (\ref{eq7}) is valid for \emph{all} operators
$O^{\text{(Semi-rdm)}}$ in the form (\ref{eq1add}), there must exist
a unitary transformation $U$ independent on the operator
$O^{\text{(Semi-rdm)}}$ (because it is constructed by a random
matrix $\sigma_1^{\text{(rdm)}}$ in subsystem \lq\lq $1$\rq\rq ~, and a random
matrix $\sigma_2^{\text{(rdm)}}$ in subsystem \lq\lq $2$\rq\rq ~), to satisfy all
$O^{\text{(Semi-rdm)}}$-orthogonal relations in the corresponding
decomposition, and such a bipartite state is separable; On the other
hand, any violation of this condition, just as we already shown, is
enough to declare the existence of quantum entanglement. In other
words, every operator $O^{\text{(Semi-rdm)}}$ in the form
(\ref{eq1add}) is a witness \cite{witness} of bipartite entanglement
under the violation of the condition (\ref{eq7}).

The singular values $\{\lambda_i\}$ of the symmetric matrix $S$
(\ref{eq6}) are equal to the square root of the eigenvalues of the
matrix $S^{\dag}S$ \cite{SVD}, and as well the square root of the
eigenvalues of the following matrix,
\begin{align} \label{eq9}
\rho\left[O^{\text{(Semi-rdm)}}\right]^{\dag}\rho^{*}O^{\text{(Semi-rdm)}},
\end{align}
which is very similar to the form used in the concurrence paper
\cite{concurrence}, and can be considered as its generalization in
high dimensional systems.

At present, we can not verify the separability of a bipartite state in high dimensions
through finite tests on the condition (\ref{eq7}), unless it is a
pure state. But we here provided an efficient way for verifying the
entanglement of a bipartite state, because any operator
$O^{\text{(Semi-rdm)}}$ in the form (\ref{eq1add}) is an
entanglement witness under the violation of the condition (\ref{eq7}). Our
numerical results show that the larger entanglement a bipartite
state contains, the fewer tests it usually requires to find a violation
by randomly generating the matrices $\sigma_1^{\text{(rdm)}}$ and
$\sigma_2^{\text{(rdm)}}$ in the form (\ref{eq1add}).


Now we come to the multipartite case. A multipartite pure state is
said to be entangled only when it cannot be written as a tensor
product of the states in the subsystems. This definition does not distinguish between
\lq\lq truly multipartite\rq\rq ~entanglement and low-partite
entanglement \cite{truemultientanglecriterion}. For example, both the Greenberger-Horne-Zeilinger state and the $W$ state in the three-qubit system are tripartite entangled states according to the above definition, but only the former one has nonzero $3$-tangle for \lq\lq truly tripartite\rq\rq ~entanglement \cite{residual}. Our following discussion on the separability criterion of multipartite states is based on this definition of entanglement, no matter what kind of entanglement, \lq\lq truly multipartite\rq\rq ~entanglement or low-partite
entanglement, is contained.

For a separable multipartite pure state,
\begin{align} \label{eq10}
|\psi>=|\psi_{1}>\otimes|\psi_{2}>\otimes
\cdots\otimes|\psi_{N}>,
\end{align}
each subsystem is separable with the rest subsystems as a whole.
The inverse situation is also true. Among a multipartite system, if
every subsystem is separable with the rest subsystems as a whole,
this multipartite pure state is separable. That is to say, multipartite
entanglement can be featured by bipartite entanglement. Thus the
entanglement degree of a multipartite pure state can be quantified
by
$C^{(N)}(|\psi>)=\sqrt{\frac{1}{2}\sum_{k=1}^N|C^{(2)}(|\psi_{k\overline{k}}>)|^2}$,
where the superscript \lq\lq$(N)$\rq\rq ~denotes $N$-partite
system, $C^{(2)}$ is the bipartite
entanglement defined in Eq.(\ref{eq2}), and $|\psi_{k\overline{k}}>$ is the bipartite version of the quantum state $|\psi>$ in Eq.(\ref{eq10}) by considering $(N-1)$ subsystems, excluding the $k$th one, as a whole.
Although some other bipartite entanglement, e.g.
$C^{(2)}(|\psi_{(12)(\overline{12})}>)$, where the first part $(12)$ is
composed of the two subsystems \lq\lq $1$\rq\rq ~and \lq\lq $2$\rq\rq, and the other part $(\overline{12})$ is composed of the rest $(N-2)$ subsystems, can also be included to define other
types of multipartite entanglement, for example in Ref.
\cite{songheshan}, the participation of these terms only affects the
evaluation of entanglement, but does not
change the separability criterion. It means we can measure the multipartite entanglement in a simpler way, where every involved bipartite
entanglement is associated with one single subsystem and the rest
as a whole. The above entanglement measure for multipartite states can also be
described in terms of operators,
\begin{align} \label{eq11}
C^{(N)}(|\psi>)=\sqrt{\frac{1}{8}\sum_{k=1}^N\sum_{i,i^{\prime}=1}^{D_1}\cdots\sum_{j,j^{\prime}=1}^{D_N}|<\psi^{*}|O_{k\overline{k}}^{(i,i^{\prime}:\cdots:j,j^{\prime})}|\psi>|^2},
\end{align}
where the symmetric operator,
\begin{align} \label{eq12}
&O_{k\overline{k}}^{(i,i^{\prime}:\cdots:j,j^{\prime})}=\sigma_1^{(i,i^{\prime})}\otimes\cdots\otimes[\sigma_k^{(m,m^{\prime})}-(\sigma_k^{(m,m^{\prime})})^T]\otimes\cdots\otimes\sigma_N^{(j,j^{\prime})}+T.c.,
\end{align}
is associated the bipartite system composed of the $k$th subsystem and all the other
subsystems as a whole, $D_k$ is the dimension degree of the $k$th subsystem and \emph{T.c.} means matrix transpose. This is generalized \emph{O concurrence} for multipartite pure states,
which, inherited from bipartite entanglement, remains invariant under
local unitary transformations. Under this measure, the
maximally-entangled $D$-dimensional $N$-partite Greenberger-Horne-Zeilinger state
$|GHZ^{(N)}>=\frac{1}{\sqrt{D}}\sum_{i=1}^{D}|i_1i_2\cdots i_N>$
has entanglement $\sqrt{N(1-\frac{1}{D})}$, and the generalized
$N$-qubit $W$ state
$|W^{(N)}>=\frac{1}{\sqrt{N}}(|100\cdots>+|010\cdots>+\cdots+|00\cdots1>)$
has entanglement $\sqrt{2(1-\frac{1}{N})}$. Similar to the bipartite
case, the entanglement measure (\ref{eq11}) has an equivalent but much simpler version,
\begin{align} \label{eq13}
C^{(N)}(|\psi>)=\sqrt{N-\sum_{k=1}^{N}\text{tr}(\rho_k^2)},
\end{align}
where $\rho_k$ stands for the reduced density matrix of the $k$th
subsystem. This measure is a generalization of the \emph{I
concurrence} \cite{Iconcurrence} for multipartite systems.

Since the present entanglement measure for multipartite systems is
based on the bipartite entanglement between single subsystems and the rest as a whole, we can
directly generalize some conclusions from bipartite case to the
present multipartite case. For example, a multipartite pure state is
separable only when it is orthogonal to its complex conjugate with
respect to \emph{all} operators in the form (\ref{eq12}), and also their extended version,
\begin{align} \label{eq14}
O_{k\overline{k}}^{\text{(Semi-rdm)}}=\sigma_1^{\text{(rdm)}}\otimes\cdots\otimes[\sigma_k^{\text{(rdm)}}-(\sigma_k^{\text{(rdm)}})^T]\otimes\cdots\otimes\sigma_N^{\text{(rdm)}}+T.c.,
\end{align}
where $\sigma_k^{\text{(rdm)}}$ is a random $D_k\times D_k$ matrix in the $k$th subsystem.

For a separable multipartite state $\rho$, there exists a decomposition, $W=\sqrt{\rho}U$, where each
\lq\lq subnormalized\rq\rq ~pure state $|W_i>$ ($i$th column vector
of $W$) is a tensor product state in the form (\ref{eq10}), thus is
$O_{k\overline{k}}^{\text{(Semi-rdm)}}$-orthogonal to its complex
conjugate. Similar to the above case, the matrix
$W^TO_{k\overline{k}}^{\text{(Semi-rdm)}}W$ has to be a hollow
matrix, which means the singular values $\{\lambda_i\}$ in
decreasing order of the symmetric matrix,
\begin{align} \label{eq15}
S=(\sqrt{\rho})^TO_{k\overline{k}}^{\text{(Semi-rdm)}}\sqrt{\rho},
\end{align}
have to satisfy the condition (\ref{eq7}). Now we conclude the
separability criterion for a mixed multipartite state, that is, the
condition (\ref{eq7}) for the singular values of the above matrix
(\ref{eq15}) holds valid for \emph{all} symmetric operators in the form
(\ref{eq14}). A violation by any operator in the form (\ref{eq14})
is enough to declare the existence of multipartite entanglement.
So every operator
$O_{k\overline{k}}^{\text{(Semi-rdm)}}$ in the form (\ref{eq14}) is a
witness of multipartite entanglement under the violation of the condition
(\ref{eq7}). This huge class of entanglement witnesses can help us efficiently verify multipartite entanglement, even for
mixed multipartite states. Here
we also note that the singular values $\{\lambda_i\}$ of the
symmetric matrix (\ref{eq15}) are the square root of the eigenvalues
of $S^{\dag}S$, and as well the square root of the eigenvalues of
the following matrix,
\begin{align} \label{eq16}
\rho\left[O_{k\overline{k}}^{\text{(Semi-rdm)}}\right]^{\dag}\rho^{*}O_{k\overline{k}}^{\text{(Semi-rdm)}},
\end{align}
which is directly determined by the density matrix
$\rho$ and an operator
$O_{k\overline{k}}^{\text{(Semi-rdm)}}$ constructed by random
matrices in the way (\ref{eq14}).

To summarize, a group of symmetric operators with only two nonzero matrix elements are introduced to carry out the separability criterion of a general quantum state. Every symmetric operator, formulated as a
tensor product of an antisymmetric operator in one subsystem
and an antisymmetric operator in other subsystems as a whole, and their arbitrary linear combinations are entanglement witnesses for multipartite quantum states. Based on these symmetric operators, Wootters' separability criterion for two-qubit states is generalized to multipartite systems in arbitrary dimensions. How to distinguish all mixed entangled states from separable ones within a finite number of tests is the next challenging question worthy of investigation. Its answer might directly bring us a computable entanglement measure for mixed multipartite states.

This work was supported by the national Natural Science Foundation
of China under Grant Nos. 11174118, 11174026 and 11264018, and the Natural Science Foundation of Jiangxi Province, China under
Grant No. 20114BAB212003. We also thank Prof. J.P. Dowling for helpful discussions.

\emph{Note added.}-After completion of this work, we became aware of
a related work by Chen, Ma, G\"{u}hne, and Severini\cite{ZHchen}.



\end{document}